\documentclass[a4paper]{jpconf}
\usepackage[cp1252]{inputenc}
\usepackage{iopams}
\begin{document}
\title{Explosive shock tube of xenon non--ideal plasma for proton radiography}

\author{N~S~Shilkin, D~S~Yuriev and V~B~Mintsev}

\address{Institute of Problems of Chemical Physics of the Russian Academy of Sciences, Academician Semenov Avenue 1, Chernogolovka, Moscow Region 142432, Russia}

\ead{shilkinns@mail.ru}

\begin{abstract}
A high explosive shock tube of non-ideal gaseous plasma for proton radiography is described. The gas dynamic flow in the shock compressed xenon at initial pressure of 7~ bar was investigated in the tube. The velocity of the shock wave in xenon and the associated particle velocity were measured by a high-speed rotating mirror streak camera. Experimental time-distance data was used for approximation of the velocities by exponential decay functions. The shock tube is intended for generation of non-ideal plasma of xenon at the pressure of $5\textrm{--}12$~kbar, the density of $0.24\textrm{--}0.3$~g/cm$^3$ when the initial pressure is about 7~bar.
\end{abstract}

\section{Introduction}
The thermodynamic properties of the plasma with strong interparticle interactions are of great interest in astrophysics, high energy density physics, planetary science, physics of non-ideal plasma and numerous practical applications \cite{Fortov-2011}. A considerable number of theoretical and experimental studies of  equation of state of non-ideal plasma ($\Gamma=E_{INT}/E_{KIN}>1,~E_{INT}$ is energy of interparticle interaction, $E_{KIN}$  is thermal energy of particles) was performed. A set of theoretical models for states with weak interparticle interactions was developed. The construction of strict and consistent theory of non-ideal plasma is still a challenging task due to the difficulties of correct accounting of strong interparticle interaction.

Strongly coupled plasma is typically generated by dynamic methods with the help of powerful drivers like high explosives, lasers, charged particle beams, x-rays, light gas guns and other pulse power devices. Shock compression of gases is one of the common techniques of generation of a non-ideal plasma. The density of shock-compressed plasma is usually defined by x-ray pulse radiography \cite{Egorov-2011} or derived from the conservation laws by the measured kinematic parameters of the gas-dynamic flow \cite{Zel}. Typical accuracy of density determination of single shock-compressed gaseous xenon by the measured velocities of shock wave and compressed gas was presented in \cite{Fortov-76}, the estimates gave the values about 7--18\%. The density of shock compressed weakly ionized xenon at a diaphragm tube was determined by a pulsed x-ray method with accuracy of 4--6\%\cite{Bushman-75}, the density of non-ideal plasma of argon in a linear explosive generator was measured with accuracy of 8 \%\cite{Bespalov-75}. Theoretical advances in thermodynamics of non-ideal plasma may be greatly supported by the precise experimental data on density that could be obtained by proton radiography \cite{King-99}. Modern proton radiography systems equipped with special magnetic optics provide high spatial resolution at the level of 17--250$~\mu$m depending on thickness of the studied object  \cite{King-99, Mottershead-03, Varentsov-16, Antipov-10} and high temporal resolution of several tens of nanoseconds. It is expected that the accuracy of determination of density of shock compressed gas may be better than 1\% if the proton microscope registering device had dynamic range about 1000 \cite{Kantsyrev-18}.

 The details on shock tube of non-ideal gaseous plasma for proton radiography at PUMA proton microscope \cite{Kantsyrev-14} at TWAC--ITEP \cite{Golubev-10} facility are described. A description of high-explosive generators for proton radiography and physical program of experiments at PUMA were reported elsewhere \cite{Kolesnikov-12, Mintsev-18}. The new experimental results on caloric equation of state of shock-compressed non-ideal plasma of xenon are also presented.

\section{Explosively driven shock tube}
Explosively driven shock tubes are actively used in the experimental studies of thermodynamic and transport properties of strongly coupled plasmas and extreme states of matter \cite{Mintsev-06}. Linear explosive generators have the simplest construction among shock tubes \cite{Mintsev-82}, a shock wave in them is formed by quasi-one-dimensional expansion of products of detonation in the gas. It was shown earlier that non-ideal plasma ($\Gamma\geq$~1) will be generated by a single shock compression and irreversible heating of the gaseous xenon with an initial pressure more than 2~bars and the speed of the shock wave in it more than 4--5~km/s  \cite{Fortov-75}. The ionizing shock wave should be formed by the expansion of detonation products of a high explosive like hexogen in the gas. 

This construction of an explosive generator was adapted to radiographic experiments at the PUMA microscope. The field of view of the microscope was accepted equal to 22--25~mm and the external diameter of the generator was made of 25~mm to ensure registering at least half of cross-section of the symmetric shock tube. The developed shock tube (figure 1) contained only 20--30~g of high explosive as the explosive chamber integrated with the proton microscope was approved to charges with mass no more than 70~g. 
\begin{figure}[t]
\centering\includegraphics[width=0.6\columnwidth]{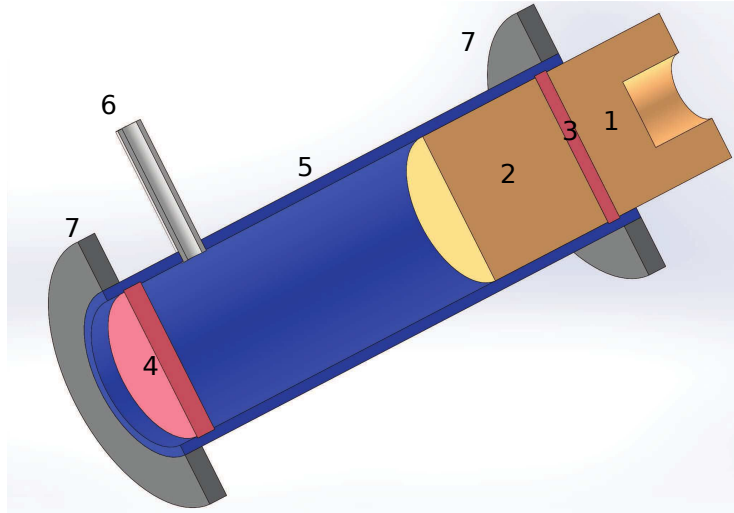}
\caption{Cross view of shock tube of non-ideal plasma for proton radiography:
1---intermediate charge with a hole for a detonator; 2---active charge; 3---flange for high explosive; 4---transparent flange; 5---shock tube channel; 6---gas inlet; 7--- fastening rings.}
\end{figure}
The shock tube channel was manufactured of transparent organic glass with the wall thickness of 2~mm. Two flanges were pressed in ring grooves made on the inner surface of the channel. Additional air-tightness of the coupling details was provided by liquid glue containing cyanoacrylate. One of the flanges was used for fastening of the main and intermediate charge of pressed desensitized hexogen. The detonation impulse was transferred through this flange from one charge to another. The second transparent flange was used for high speed optical filming of the shock-wave processes in the generator. A copper pipe with the external diameter of 4~mm and the length of several cm was glued in the body of the tube, it was used as a connection to a vacuum-gas system. The diameter of the active charge was less than the inner diameter of the tube by 2~mm to organize the channel effect \cite{Woodhead-47} in the gas gap. This technique flattens the detonation front without application of an explosive lens. The gas gap between the charge and the wall delayed the arising of a shock wave precursor along the tube wall. The precursor wave was described for the first time in \cite{Shreffler-54} but still there is no strict and self-consistent explanation of this phenomenon. The shock tube could be equipped by fastening rings which were placed on the external surface of the channel. The described shock tube was designed as a generator of shock-compressed inert gases at initial pressure up to 12~bar and the initial temperature close to normal. 
\section{Plasma diagnostic and results}
A fast rotating mirror camera was used to record the light emitted from the shock-compressed xenon plasma moving in the shock tube. The camera slit was oriented parallel to the shock tube axis. The slit had a slightly V-shaped form to expand the dynamic range of optical measurements, so the upper part of the image is brighter than the lower one. A mirror inclined at an angle of 45~degrees to the flange was placed at a small distance from its surface. This geometry allowed recording a light radiation from plasma in two orthogonal directions: from the top of the tube and from the direction orthogonal to the surface of the tube. The image of the slit was recorded on a high resolution black and white film. Initial pressure in the tube was measured by a standard pressure gauge taking into account the local atmosphere pressure of 746~mm~Hg. Four experiments with gaseous xenon of purity $5.5$ at initial temperature about 293~K and pressure about 7~bar were conducted. Only the length between the free surface of the charge and the transparent flange was varied. A typical inverted streak photography of shock compressed xenon plasma moving inside the shock tube is presented in figure 2. 
\begin{figure}[t]
\centering\includegraphics[width=0.6\columnwidth]{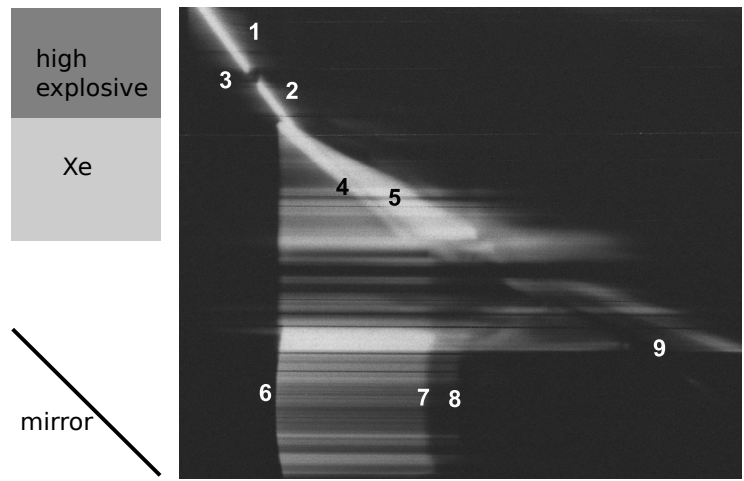}
\caption{Streak image of shock-wave processes in the tube with xenon at initial pressure of 7~bar:
1---channel effect before marker; 2---channel effect after marker; 3---opaque marker; 4---shock compressed xenon from the gap and wall shock wave precursor; 5---shock-compressed plasma behind flat shock wave and products of detonation; 6---shock wave exit on the free surface of the charge; 7---beginning of reflection of shock wave front from the flange; 8---back side of detonation products arrived at the flange; 9---interaction of shock wave in air with the mirror.}
\end{figure}
The image of the slit moved on a 35~mm film from left to right with the speed of $2.23$~km/s. The top of the picture is recorded through the surface of the shock tube. The bright white line  was induced by a luminescence of the multiply shock-compressed gas between the channel and the charge. A dark spot separated the white line in two parts, it was caused by absorption of the radiation by a ring opaque mark on the surface of the shock tube. This mark was used for assignment a reference length in the tube. A light emission of the plasma slug formed in the gap between the charge and the shock tube wall, the wall shock wave precursor, shock compressed xenon plasma is registered at the later moments of time. The bottom part of the image is formed by the light reflected from the mirror. A light output appeared in the whole cross-section of the tube when the ionizing shock wave was generated in the gas after the detonation wave arrived at the free surface of the charge. The intensity of the radiation increased quickly during 60--120~ns with the growth of the optical thickness of the shock-compressed plasma in the beginning of the record. It decreased a little over time later because the shock wave slowed down. The arrival of the products of detonation to the free surface of the flange was determined by noticeable falling of the light radiation recorded through the mirror. This moment was accepted as the beginning of the reflection. The velocities of the shock-compressed plasma and the products of detonation were accepted to be equal on a contact surface. An expansion of the external face of the flange in the air happened later. The registration come to the end when products of expansion interact with the mirror. The time dependence of the light emission from the fixed area of the slit is shown in figure 3.
\begin{figure}[t]
\centering\includegraphics[width=0.5\columnwidth]{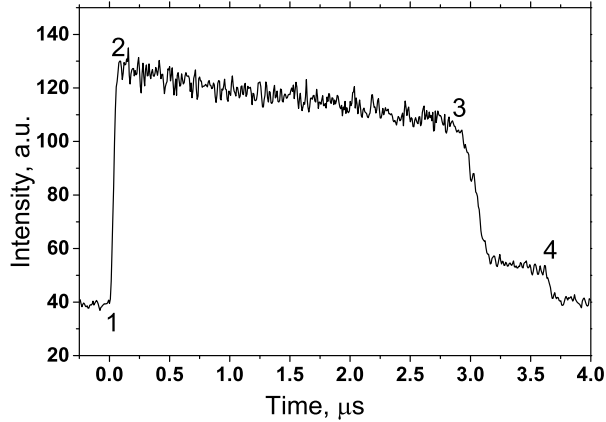}
\caption{Time dependence of the light emission out of a shock-compressed xenon plasma: 
1---shock wave exit on the free surface of a charge; 2---end of increase of optical thickness of shock compressed plasma; 3---beginning of reflection of shock wave front from the flange; 4---back side of detonation products arrived at the flange.}
\end{figure}
Propagation of a planar shock wave in a cylindrical shock tube was studied by many researchers. It was proposed to fit the distance-time data of a shock wave position in the channel by an exponential dependence in \cite{Davis-06}: 
\begin{equation}\label{XTD}
x(t)=p_\mathrm{1}\bigg(1-\exp\bigg(-\frac{t}{p_\mathrm{2}}\bigg)\bigg),
\end{equation}
where $x$ is a distance between the free surface of a charge and an observed plane in mm, $t$ is time of shock-wave propagation to the observed plane in $\mu$s, $p_{1}$ and $p_{2}$ are constants that were calculated according to the experimental data. Such type of approximation was used for simplicity. Differentiating expression (\ref{XTD}), one could receive that the velocity of the shock wave $D(t)$ is decreased by the exponential law: 
\begin{equation}\label{DT}
D(t)=\frac{p_\mathrm{1}}{p_\mathrm{2}}\exp\bigg(-\frac{t}{p_\mathrm{2}}\bigg)=D_\mathrm{0}\exp\bigg(-\frac{t}{p_\mathrm{2}}\bigg),
\end{equation}
where $D_{0}=p_{1}/p_{2}$ is expressed in mm/$\mu$s and $p_{2}$ in $\mu$s.

We applied this type of fitting not only for approximation of the speed of the shock wave front but also for the approximation of the speed of shock-compressed plasma $U(t)$ behind the incident shock wave: 
\begin{equation}\label{XTU}
x(t)=p_\mathrm{3}\bigg(1-\exp\bigg(-\frac{t}{p_\mathrm{4}}\bigg)\bigg),
\end{equation}
\begin{equation}\label{UT}
U(t)=\frac{p_\mathrm{3}}{p_\mathrm{4}}\exp\bigg(-\frac{t}{p_\mathrm{4}}\bigg)=U_\mathrm{0}\exp\bigg(-\frac{t}{p_\mathrm{4}}\bigg),
\end{equation}
where $U_{0}=p_{3}/p_{4}$ is expressed in mm/$\mu$s and $p_{4}$ in $\mu$s.

Parameters $p_{1}$, $p_{2}$, $p_{3}$ and $p_{4}$ were calculated from the data on time-distance of shock wave front and shock compressed plasma of xenon in the conducted experiments (table 1). The following values were received: $p_{1}=64.2$~mm, $p_{2}=10.6~\mu$s, $p_{3}=58.7$~mm, $p_{4}=11.6~\mu$s. Standard error of $p_{1}$ and $p_{2}$ approximation was of $6.5$~mm, standard error of $p_{3}$ and $p_{4}$ approximation was of 5--6~$\mu$s. So the relative accuracy of the approximated parameters $p_{1}\textrm{--}p_{4}$ was about 10\%.

\Table{\label{Table_1}Time-distance data of shock wave front and shock compressed plasma.}
\br
Length of the tube with gaseous xenon,~mm  & 15.7& 9.7&6.7&4.5\\
Time of propagation of shock wave front,~$\mu$s & 2.98&1.73& 1.19&0.76 \\
Time of propagation of shock-compressed plasma,~$\mu$s & 5.12 & 4.69& 4.49&4.30\\
Initial pressure in xenon,~bar & 7.00 & 7.04 & 7.08&7.00\\
 \br
\endTable
The dependences of speed of the front $D(t)$ and speed of plasma $U(t)$ on time are shown in figure 4.
\begin{figure}[t]
\centering\includegraphics[width=0.5\columnwidth]{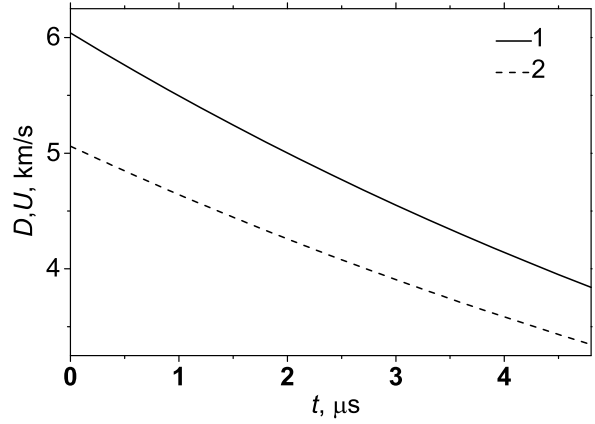}
\caption{Time dependent approximation of $D(t)$ and $U(t)$: 1---velocity of shock wave front; 2---velocity of shock compressed plasma.}
\end{figure}
The obtained results were presented in $D\textrm{--}U$ coordinates in figure 5, where results from other studies \cite{Gryaznov-80, Fortov-76} were plotted also. Left part of the obtained curve 4 coincided with earlier obtained results. The right part of the curve 4  had a little difference with points in the right upper corner. 
Finally pressure, density and change in specific internal energy were estimated by the  well known conservation laws on shock discontinuity\cite{Zel}. New point are also presented in coordinates pressure-specific volume (figure 6). The initial xenon thermodynamic parameters were calculated from the measured pressure, temperature and reference data \cite{Vargaftik-72}. Non-ideal plasma of xenon with the following thermodynamic parameters were generated in the described shock tube: pressure of 5--12~Kbar, density of $0.24\textrm{--}0.30$~g/cm$^3$, specific internal energy of $5.6\textrm{--}13$~MJ/kg. A  thermal equation of state could be used for an estimation of plasma temperature. The chemical model with Coulomb interactions considered in frames of the Debye approximation in a grand canonical ensemble \cite{Gryaznov-73} gave the plasma temperature of 20--30~kK and the value of Coulomb non-ideality parameter about $\Gamma\sim2.5$. The similar approach was used earlier for the description of thermodynamic properties of shock compressed plasma of xenon at  $\Gamma$ up to 3 \cite{Shilkin1-03, Shilkin2-03}. 
\begin{figure}[t]
\centering\includegraphics[width=0.5\columnwidth]{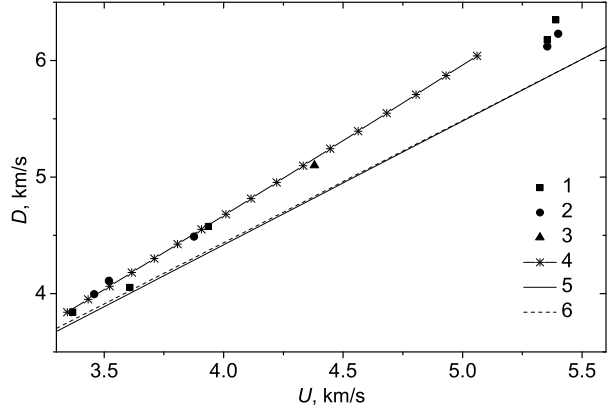}
\caption{Hugoniot of xenon in $D\textrm{--}U$ coordinates: 1---experiment at initial pressure 5~bar\cite{Gryaznov-80}; 2---experiment at initial pressure 10~bar\cite{Gryaznov-80}; 3---experiment at initial pressure 10~bar \cite{Fortov-76}; 4---approximation of experimenatal data in present work at initial pressure 7~bar; 5---Debye approximation in a grand canonical ensemble of shock-compressed xenon at initial pressure 5~bar\cite{Gryaznov-80}; 6---Debye approximation in a grand canonical ensemble of shock-compressed xenon at initial pressure 10~bar\cite{Gryaznov-80}.}
\end{figure}
\begin{figure}[t]
\centering\includegraphics[width=0.55\columnwidth]{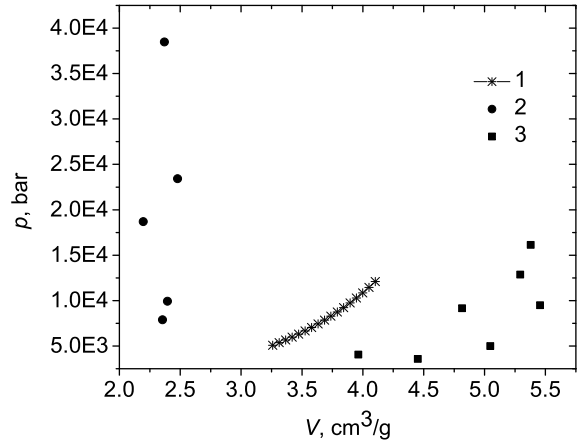}
\caption{Hugoniot of xenon in coordinates pressure--specific volume: 1---approximation of experimenatal data in present work at initial pressure 7~bar; 2---experiment at initial pressure 10~bar\cite{Gryaznov-80}; 3---experiment at initial pressure 5~bar\cite{Gryaznov-80}.}
\end{figure}
\section{Conclusions}
A high explosive shock tube of a non-ideal gaseous plasma for proton radiography was developed for proton radiography. It contained only 20~g of high explosive and had the diameter of 25~mm. It was shown that non-ideal plasma with $\Gamma\sim2.5$ will be generated under shock loading of xenon at the initial pressure of 7~bar in this shock tube. The generator could be used for investigation of equation of state at proton microscope PRIOR at the GSI Helmholtzzentrum fur Schwerionenforschung (Darmstadt, Germany) and at the designed radiographic setup at the Institute for Nuclear Research (Troitsk, Russia).
\ack
This work was supported by the Program of the Presidium of the Russian Academy of Sciences, the experiments were carried out with the use of equipment of Interregional Explosive Center for Collective Use and Unique Scientific Facility Explosive Test Installation.

\section*{References}
\bibliographystyle{iopart-num}
\bibliography{2018tubes01}

\end{document}